\documentclass[12pt,aps,prd,a4paper,superscriptaddress,nofootinbib,onecolumn]{revtex4}

\usepackage{amsfonts}
\usepackage{amsmath}
\usepackage{babel}
\usepackage[active]{srcltx}
\usepackage{graphicx}
\usepackage{color}
\usepackage{float}
\usepackage{changebar}
\usepackage{esint}
\usepackage{dsfont}
\usepackage{multirow}
\usepackage[usenames,dvipsnames]{xcolor}
\usepackage[colorlinks=true,pdfstartview=FitV,linkcolor=blue,citecolor=blue,urlcolor=blue,breaklinks=true]{hyperref}
\usepackage{braket}
\usepackage{mathtools}
\usepackage{slashed}
\usepackage{ulem}
\usepackage{empheq}
\usepackage{multirow}
\usepackage[utf8]{inputenc}
\usepackage[T1]{fontenc}

\makeatletter
\RequirePackage{color}

\begin{document}

\title{Kinks and double-kinks in generalized $\phi^{4}$- and $\phi^{8}$-models}

\author{F. C. E. Lima}
\email{fcelima.fisica@gmail.com (F. C. E. Lima)}
\affiliation{Programa de P\'{o}s-gradua\c{c}\~{a}o em F\'{i}sica, Universidade Federal do Maranh\~{a}o, Campus Universit\'{a}rio do Bacanga, S\~{a}o Lu\'{i}s (MA), 65080-805, Brazil.}

\author{R. Casana}
\email{rodolfo.casana@gmail.com (R. Casana)}
\affiliation{Programa de P\'{o}s-gradua\c{c}\~{a}o em F\'{i}sica, Universidade Federal do Maranh\~{a}o, Campus Universit\'{a}rio do Bacanga, S\~{a}o Lu\'{i}s (MA), 65080-805, Brazil.}
\affiliation{Departamento de F\'{i}sica, Universidade Federal do Maranh\~{a}o, Campus Universit\'{a}rio do Bacanga, S\~{a}o Lu\'{i}s (MA), 65080-805, Brazil.} 

\author{C. A. S. Almeida}
\email{carlos@fisica.ufc.br (C. A. S. Almeida)}
\affiliation{Universidade Federal do Cear\'{a} (UFC), Departamento de F\'{i}sica, Campus do Pici, Fortaleza - CE,  60455-760 - Brazil.}
\affiliation{Departament de F\'isica Te\`orica and IFIC, Centro Mixto Universitat de
Val\`{e}ncia - CSIC. Universitat de Val\`{e}ncia, Burjassot-46100, Valencia, Spain.}
\vspace{1cm}

\begin{abstract}
\vspace{0.5cm}
\noindent \textbf{Abstract:} Examining the $\phi^{4}$ and $\phi^{8}$ models within a two-dimensional framework in the flat spacetime and embracing a theory with unconventional kinetic terms, one investigates the emergence of kinks/antikinks and double-kinks/antikinks. We devote our study to obtaining the field configurations with minimal energy, i.e., solutions possessing a Bogomol'nyi-Prasad-Sommerfield's bound. Next, to accomplish our goal, we adopt non-polynomial generalizing functions, namely, hyperbolic sine and cosine functions: the first produce BPS potentials exhibiting a minimum at $\phi=0$, facilitating the emergence of genuine double-kink-type configurations. Conversely, the second promotes the rise of kink-type solutions.
\end{abstract}
\maketitle

\thispagestyle{empty}

\section{Introduction}
In the quantum field theory framework, specifically in nonlinear models, extended objects with conserved topological charges arise, and among these structures are the so-called ``kinks'' or domain walls; generally speaking, kinks provide a covariant description of extended particles \cite{Finkelstein,Vachaspati}. In such a context, several works in the literature describe kinks emerging inside different theories \cite{Graham,Zhong,Fabiano}. Pioneering studies on kinks belong to Finkelstein \cite{Finkelstein}, which aimed to establish the properties that a field theory must possess to allow the existence of kinks and under what circumstances these structures can have spin $1/2$. Typically, kinks are solitons emerging as nonperturbative solutions in some field theories or due to particle interactions \cite{Albayrak,Rajaraman,Manton,Vilenkin}. One highlights that the literature dedicates significant attention to studying the nonperturbative topological structures and their interactions with perturbative excitations. Besides, one has also directed substantial attention towards static and dynamic research of single domain walls. Such emphasis is justified by the challenge posed for any attempt to develop theories, even in the static regime, describing multiple domain walls (i.e., double kinks) while preserving the $\mathbb{Z}_2$ symmetry in flat spacetime. In this study, we consider the two-dimensional flat spacetime, and maintaining the $\mathbb{Z}_2$ symmetry intact, we seek to produce topological structures described by double domain walls (double-kinks).

Exciting models for investigating domain walls are higher-order theories such as, for example, the $\phi^6, \phi^8, \phi^{10}, \dots$ models \cite{Zhong2}. The interest in these theories stems from the ability of solitons (kinks, double-kinks, or multi-kinks) to experience long-range interactions \cite{Belendryasova,Mello,Khare}. Within this scenario, the emergence of structures can reveal characteristics of notable relevance. Such a phenomenon arises because of the long-range forces, experimented by the kinks in interaction, that can potentially deform the topological structures associated with these higher-order configurations \cite{Zhong2}. One justifies the interest in obtaining double-kink-like structures by their various practical applications, such as the research on the emergence of multi-kinks \cite{Saadatmand,Marjaneh} and the study of the collisions involving multi-kinks \cite{Gani2}.

Striving to accomplish our purpose, we will employ the BPS formalism, which arises in field theory as a convenient tool for solving the equations of motion of static topological fields, allowing us the reduction of the order of the Euler-Lagrange equation \cite{Bogomolnyi,PS}. Concisely, BPS solutions characterize the field configurations with the lowest energy in the system. Although the BPS formalism is particularly useful in describing canonical theories, it is feasible also to apply it to some non-canonical theories and ensure the presence of the BPS property  \cite{Acalapati}. Furthermore, one finds in the literature that models with BPS properties confer stability to the static model and imply stressless configurations  (i.e., with $T_{ij}=0$) \cite{Bazeia0}. Additionally, the BPS formalism has become essential in describing BPS defects with non-canonical kinetic terms \cite{FLima1,FLima2,Atmaja}.

In light of these motivations and adopting the BPS approach, the purpose is to study the emergence of kinks and double-kinks with compact-like features through a generalized scalar theory. To accomplish this purpose, the $\phi^{4}$ and $\phi^{8}$ models generalized by non-polynomial functions, specifically hyperbolic sine and cosine functions, will be analyzed.
Furthermore, we highlight that the chosen generalizing functions preserve the $\mathds{Z}_2$ symmetry original to both models. Remarkably, the hyperbolic sine enables the BPS potential to attain a minimum at $\phi=0$, promoting the emergence of double-kink type configurations.

We outline our manuscript into four sections. In Sec. \ref{SecII}, we discuss the theory of generalized scalar field and impose a constraint on the potential such that the generalized dynamics model admits the BPS property.  In Sec. \ref{SecIII}, we study four cases, demonstrating for the generalized dynamics with hyperbolic functions, the models ($\phi^{4}$ and $\phi^{8}$) with $\mathbb{Z}_2$ symmetry can engender either kink or double-kink type configurations. Additionally, one announces our findings in section \ref{SecIV}.

\section{The BPS scalar field generalized theory}\label{SecII}

We start our study by investigating a mechanism that can facilitate the production of double-kink structures in a flat two-dimensional theory\footnote{We use the metric signature $\eta^{\mu\nu}=(+,-)$.}. Thus, one proposes a generalized non-canonical theory with $\mathds{Z}_2$ symmetry, described by the action\footnote{Naturally, one can track the non-canonical theory to a canonical theory by parameterizing the scalar field. For further details, see Appendix A.}
\begin{align}\label{Eq2}
    S=\int\, dx^2\, \bigg[\frac{1}{2}f(\phi)\,\partial^\mu\phi\,\partial_\mu\phi-V(\phi)\bigg],
\end{align}
where $\phi$ denotes the real scalar field, $f(\phi)$ will be a generalizing function, and $V(\phi)$ represents the potential. Moreover, the literature highlights the ability to produce new effects capable of simulating geometric modifications that impact the physical aspects of topological structures through the extension of the theory's symmetry \cite{Bazeia}. Generally speaking, to generate contracted geometrically structures or deformed kinks because of the emergence of internal structures, one adopts scalar theories with an extended symmetry, for instance, the action  $\mathds{Z}_2\times\mathds{Z}_2$ symmetric, namely,
\begin{align}\label{Eq3}
    S=\int\, d^2 x\, \bigg[\frac{1}{2}f(\xi)\partial^\mu\phi\partial_\mu\phi+\frac{1}{2}\partial^\mu\xi\, \partial_\mu\xi-V(\phi,\,\xi)\bigg].
\end{align}

By regarding the action (\ref{Eq2}), one obtains the following field equation
\begin{align}\label{Eq4}
    f\,\partial^\mu\partial_\mu\phi+\frac{f_\phi}{2}\partial^\mu\phi\,\partial_\mu\phi+V_\phi=0,
\end{align}
where one defines $f_\phi=\frac{\partial f}{\partial\phi}$ and $V_\phi=\frac{\partial V}{\partial\phi}$.

Let us now proceed to particularize the theory for the static case, i.e., choosing $\phi\equiv \phi(x)$, where $x$ denotes the position. This consideration enables us to express the equation of motion as,
\begin{align}\label{Eq5}
    f\frac{d^2\phi}{dx^2}+\frac{1}{2}f_\phi\, \bigg(\frac{d\phi}{dx}\bigg)^2-V_\phi=0.
\end{align}

To implement the BPS technique, we write the static $\phi(x)$-field energy, which reads as
\begin{align}\label{Eq6}
    \mathrm{E}=\int_{-\infty}^{\infty} \,\bigg[\frac{f(\phi)}{2}\bigg(\frac{d\phi}{dx}\bigg)^2+V(\phi)\bigg]dx.
\end{align}
Now, we establish the boundary conditions to be satisfied in the model to ensure finite-energy configurations. For this, the integrands in  (\ref{Eq6})  must obey, respectively,
\begin{equation}
\lim_{|x|\rightarrow\infty} \sqrt{f(\phi)}\;\frac{d\phi}{dx} =0, \quad\mbox{and}\quad \lim_{|x|\rightarrow\infty}  V(\phi) =0. \label{BCEn}
\end{equation}

Allow us to rearrange the field energy as
\begin{align}\label{Eq7}
    \mathrm{E}=\int_{-\infty}^{\infty} \bigg[\frac{1}{2f} \bigg(f\frac{d\phi}{dx}\mp W_\phi\bigg)^2+V(\phi)-\frac{W_{\phi}^2}{2f} \pm\frac{dW}{dx}\bigg]dx.
\end{align}
The function $W(\phi)$ is designated as an auxiliary function, referred to as the superpotential. The choice of the function $W(\phi)$ is essential once it establishes a correspondence between $W(\phi)$ and the potential $V(\phi)$, and, therefore, with the energy \cite{Vachaspati}. Thereby, let us consider that the superpotential relates to the potential through a constraint,
\begin{align}\label{Eq8}
    V(\phi)=\frac{W_\phi^2}{2f}.
\end{align}
Then, from the potential's vacuum condition in Eq. (\ref{BCEn}), we can set the boundary conditions satisfied by the superpotential as
\begin{equation}
\lim_{x\rightarrow\pm\infty} W(\phi) =W(\phi_{\pm\infty}).\label{BCSPot}
\end{equation}

Note that the $\phi(x)$-field energy will be
\begin{align}\label{Eq9}
    \mathrm{E}=\pm\int_{-\infty}^{\infty}\,\frac{dW}{dx}\, dx\,+\,\int_{-\infty}^{\infty}\,\frac{1}{2f}\bigg(f\frac{d\phi}{dx}\mp W_\phi\bigg)^2\, dx.
\end{align}

One defines the BPS energy as
\begin{align}\label{Eq10}
    \textrm{E}_{\textrm{BPS}}=\pm\int_{-\infty}^{\infty}\, \frac{dW}{dx}\,dx= \pm[W(\phi_{+\infty})-W(\phi_{-\infty})]>0.
\end{align}
Hence, the full energy  is
\begin{align}\label{Eq11}
    \textrm{E}=\text{E}_{\textrm{BPS}}+\int_{-\infty}^{\infty}\, \frac{1}{2f}\bigg(f\frac{d\phi}{dx}\mp W_\phi\bigg)^2 dx.
\end{align}
Therefore, one can infer that $\mathrm{E} \geq \textrm{E}_{\textrm{BPS}}$. Thus, at the energy bound, i.e., $\mathrm{E}=\textrm{E}_{\textrm{BPS}}$, the self-dual equation is
\begin{align}\label{Eq12}
    \frac{d\phi(x)}{dx}=\pm \frac{W_\phi}{f(\phi)}.
\end{align}
We can quickly verify that it reproduces the Euler-Lagrange equation given by Eq. (\ref{Eq5}) by considering the self-dual potential (\ref{Eq8}). Furthermore, it is essential to emphasize that another widely used approach to deforming field configurations involves introducing impurities\footnote{In this framework, the impurities refer to perturbations in the system that can affect the stability and properties of the solutions. These impurities may encompass variations in boundary conditions, discontinuities in the medium, or the presence of other solitons and their excitations.} into a theory that already describes topological structures. Such impurities allow the construction or engendering of new topological solutions. For further details, see Refs. \cite{CAdam11,CAdam22,DBazeia111}. In particular, we consider the impurities approach exposed in Ref. \cite{Slawinska}; one notes that the BPS equation (4) in there resembles our \eqref{Eq12}. However, one cannot find a generalizing function $f(\phi)$ capable of reproducing the same Euler-Lagrange equation (7) in \cite{Slawinska}. Thus, despite the impurity function $\sigma(x)$ plays a similar role to $f(\phi)$, both models do not match because the last one depends explicitly on the field $\phi$ while the impurity does not. Consequently, the BPS equation found here represents a new one able to engender new self-dual solutions.

In addition, the BPS energy density will be
\begin{align}\label{Eq13}
    \mathcal{E}_{\textrm{BPS}}=\frac{W_{\phi}^{2}}{f(\phi)}.
\end{align}

In the following section, we will choose two superpotentials engendering the $\phi^{8}$ and $\phi^{4}$ models, both possessing the $\mathds{Z}_2$ symmetry. Both models become generalized through non-polynomial functions that keep the original symmetry. We have specifically selected the hyperbolic sine and cosine functions: the hyperbolic cosine engenders kink solutions, whereas the hyperbolic sine enables the BPS potential to attain a minimum at $\phi=0$, promoting the emergence of double-kink type configurations.


\section{BPS solitons for the generalized $\phi^{8}$- and $\phi^{4}$-models\label{SecIII}}

At this juncture, one needs to select a superpotential to ensure the existence of BPS configurations. For our purpose, we will examine two distinct superpotentials: the first generating a $\phi^8$ model is defined by
\begin{align}\label{Eq14}
   W(\phi)=\sqrt{\lambda}\bigg(\nu^{4}\phi-\frac{2\nu^{2}\phi^{3}}{3}+\frac{\phi^{5}}{5}\bigg),
\end{align}
and the second engendering the $\phi^4$ model is read as
\begin{align}\label{Eq15}
    W(\phi)=\sqrt{\lambda}\bigg(\nu^{2}\phi-\frac{\phi^{3}}{3}\bigg).
\end{align}

Considering Eq. (\ref{Eq8}), the corresponding BPS potentials for the generalized $\phi^{8}$ and $\phi^{4}$ models are described as
\begin{align}\label{Eq16}
    V(\phi)=\frac{\lambda(\nu^{2}-\phi^{2})^4}{2f(\phi)},
\end{align}
and
\begin{align}\label{Eq17}
    V(\phi)=\frac{\lambda(\nu^{2}-\phi^{2})^2}{2f(\phi)},
\end{align}
respectively. These potentials are remarkable as they enable the geometric contraction of topological structures. For instance, in Ref \cite{Bazeia}, the $\phi^{4}$ theory ruled by the potential Eq. (\ref{Eq17}) has proven effective in compactifying structures engendered by the real scalar field. Similarly, in Ref. \cite{Lima}, the $\phi^{8}$ theory described by the potential (\ref{Eq16}) has been utilized to derive compact-like vortex configurations. Besides, some researchers adopt a few high-order theories responsible for describing kinks. For instance, Ref. \cite{Demirkaya} investigates the dynamics of kinks in a $\phi^6$ system. Additionally, Ref. \cite{Gani} analyzes the interaction of kinks in a $\phi^8$ theory. However, all results from these theories yield kink-like configurations or multi-kinks interpolating between two topological sectors.

As mentioned previously, we aim to obtain both kink-type solutions as genuine double-kink configurations, which can exhibit a compact-like profile. To accomplish this purpose, we will assume three generalized functions, i.e.,
\begin{align}
    f(\phi)&=\textrm{sech}^2(m^2\phi), \label{Eq18a}\\[0.2cm] f(\phi)&=\left[\sinh(m\phi^2)\right]^{-\frac{1}{2z+1}}, \label{Eq18b}\\[0.2cm]
    f(\phi)&=\left[\sinh(m\phi^{2z})\right]^{-\frac{1}{2z+1}}. \label{Eq18c}
\end{align}

One highlights that using hyperbolic potentials engendering kink configurations appears relevant in studying ferroelectric and ferromagnetic materials. This conjecture emerges because models described by polynomial- and Sine-Gordon-like interactions tend to suffer from a drastic weakness due to the rigidity of some ferroelectric materials \cite{DBaz}, so to bypass these limitations hyperbolic extensions were proposed in Refs.\cite{Kofane1,Kofane2}. Similarly, hyperbolic extensions of Calogero's model \cite{Calogero} are integrable and generate multi-soliton solutions still in confined potentials \cite{Gon}. Likewise, we highlight that the literature extensively has utilized the hyperbolic potentials in the research of supersymmetric theories $\mathcal{N}=2$ and $\mathcal{N}=4$ \cite{Fedoruk}, hairy black holes \cite{QWen}, and quintessential inflation \cite{Agarwal}. Based on this understanding, we employ a class of hyperbolic BPS potentials to investigate the new classes of topological solutions that these models support.


\subsection{$\phi^{8}$ model: the case $f(\phi)=\textrm{sech}^2(m^2\phi)$}\label{SecIIa}

\begin{figure}[!ht]
\centering
\includegraphics[height=7cm,width=8cm]{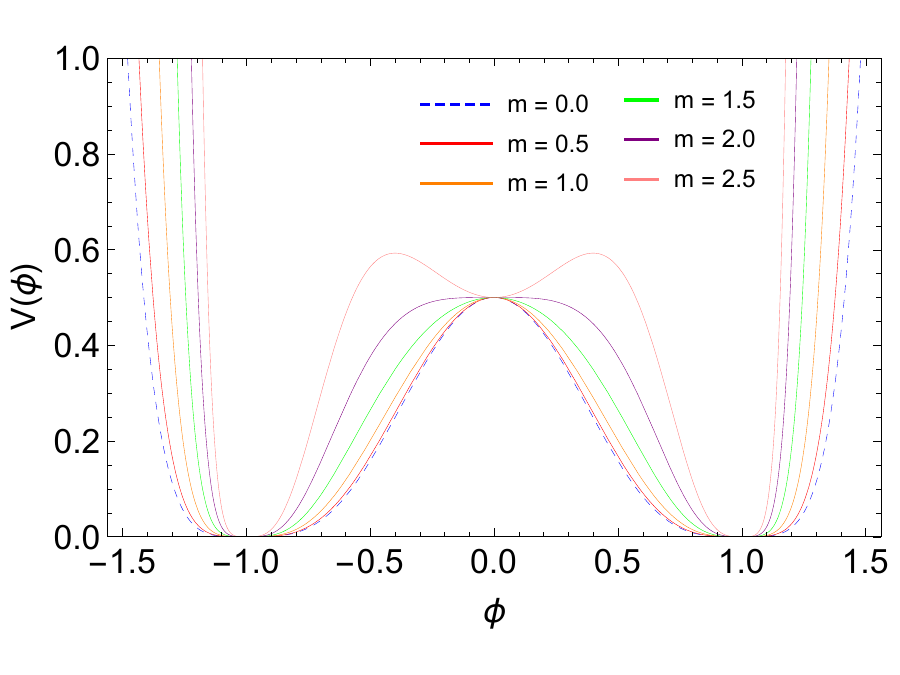} %
\caption{Potential $V(\phi)$ vs. $\phi$ [Eq. (\ref{Eq19})] with the parameter $m$ varying and keeping $\lambda=\nu=1$.} \label{fig1}
\end{figure}

Regarding the $\phi^{8}$ model, our first choice for the generalizing function is the one defined in Eq. (\ref{Eq18a}), then the BPS potential (\ref{Eq16}) becomes
\begin{align}\label{Eq19}
    V(\phi)=\frac{\lambda}{2} (\nu^{2}-\phi^{2})^4\cosh^2(m^2\phi).
\end{align}
We display the behavior of the potential (\ref{Eq19}) in Fig. \ref{fig1}. In Ref. \cite{Gani}, a $\phi^8$ model was used to investigate the formation of exotic final states in multi-kink collisions. Here, we will show that one can obtain kinks that acquire a compact-like format for the BPS theory with potential (\ref{Eq19}).

Considering the superpotential (\ref{Eq14}) in Eq. (\ref{Eq12}), one arrives at the self-dual equation
\begin{align}\label{Eq20}
    \frac{d\phi}{dx}=\pm\sqrt{\lambda}(\nu^{2}-\phi^{2})^2\cosh^2(m^2\phi).
\end{align}
In this scenario, the parameter $\lambda$ adjusts the dimension of the theory, and $\nu>0$ provides the potential's minima at $\phi=\pm v$. Here, $0\leq m<+\infty$.

To ensure the existence of a kink-like scalar field kink-like (and antikink-like), one must respect the topological boundary condition, i.e.,
\begin{align}\label{Eq21}
    \lim_{x\to\pm\infty}\phi(x)=\pm \nu,\quad \mbox{and}\quad\lim_{x\to\mp\infty}\phi(x)=\mp \nu,
\end{align}
for kinks and anti-kinks, respectively. We point out that both are compatible with the finite energy conditions established in Eq. (\ref{BCEn}).

We now present the behavior of the $\phi$ profiles near the vacuum values and around the origin. This way, we obtain,
\begin{eqnarray}
\phi \left( x\right)  &\approx &\mp v\pm \frac{1}{4v^{2}\sqrt{\lambda }\cosh
^{2}\left( m^{2}v\right) }\frac{1}{\left\vert x\right\vert },\quad
x\rightarrow -\infty, \quad \\[0.2cm]
\phi \left( x\right)  &\approx &\pm v^{4}\sqrt{\lambda }x, \quad
x\rightarrow 0,  \\[0.2cm]
\phi \left( x\right)  &\approx &\pm v\mp \frac{1}{4v^{2}\sqrt{\lambda }\cosh
^{2}\left( m^{2}v\right) }\frac{1}{\left\vert x\right\vert },\quad
x\rightarrow +\infty.
\end{eqnarray}
We first note that the profiles have a tail decaying as $|x|^{-1}$ typical of  $\phi^8$ solitons; besides, the behavior at $x=0$ confirms that we have obtained genuine kinks/anti-kinks configurations.

Adopting the topological boundaries in Eq. (\ref{Eq21}), one obtains the BPS energy, namely,
\begin{align}\label{Eq23}
    \textrm{E}_{\textrm{BPS}}=\pm \frac{16}{15}\nu^{5}\sqrt{\lambda},
\end{align}
where the BPS energy density is
\begin{align}\label{Eq24}
    \mathcal{E}_{\textrm{BPS}}=\lambda (\nu^{2}-\phi^{2})^4\cosh^2(m^2 \phi).
\end{align}

\begin{figure}[!ht]
\centering
\includegraphics[height=6cm,width=7cm]{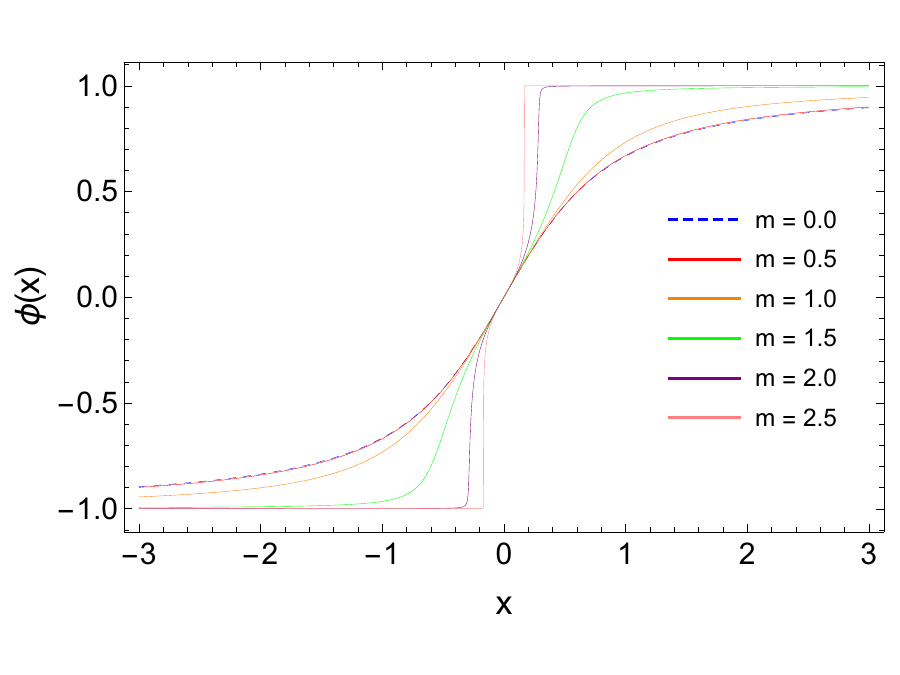} %
\includegraphics[height=6cm,width=7cm]{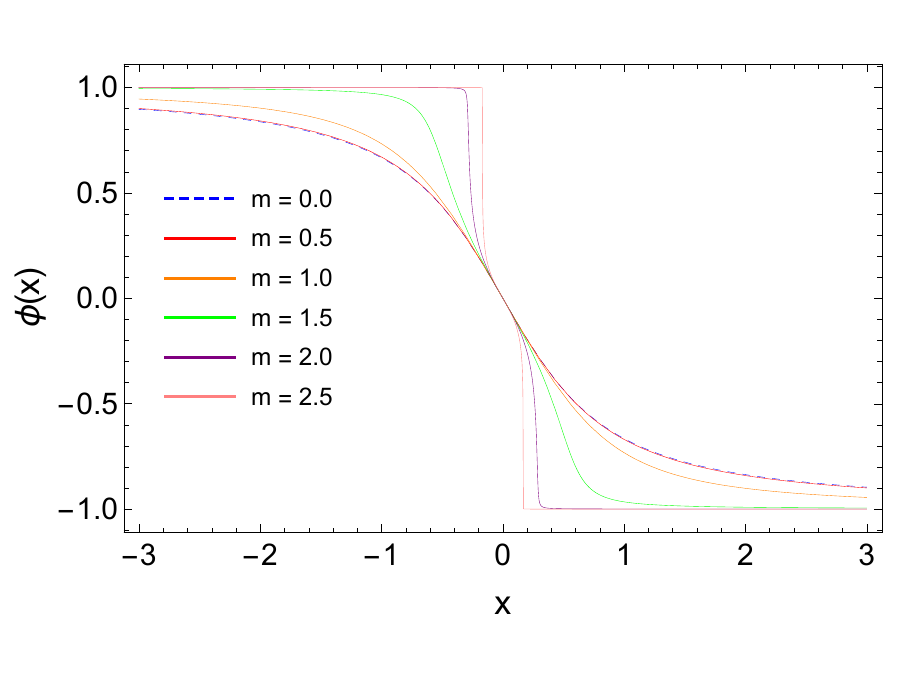}\vspace{-0.2cm}

\hspace{0.8cm} (a) \hspace{6.5cm} (b) \hfill
\caption{Plots $\phi(x)$ vs. $x$ keeping $\lambda=\nu=1$. (a) Deformed kink-like solution. (b) Deformed antikink-like solution.} \label{fig2}
\end{figure}

Let us now investigate the numerical solution of the system's self-dual equations. For this purpose, we will discretize the equation by considering the domain of the independent variable $x$ at $N$ points. For instance, $x_0, x_1, x_2, \dots, x_N$; subsequently, we will employ the interpolation method to estimate the solution of the scalar field, considering the intermediate points \cite{Hildebrand}. Adopting this methodology, the numerical solutions of equation (\ref{Eq20}) are obtained. Please see Figs. \ref{fig2}(a) and \ref{fig2}(b) for kink-like and antikink-like configurations, respectively. For the sake of simplicity, one assumes $\lambda=\nu=1$. This choice does not entail losses and facilitates our analysis once these parameters only adjust the field amplitudes.

\begin{figure}[!ht]
\centering
\includegraphics[height=6cm,width=7cm]{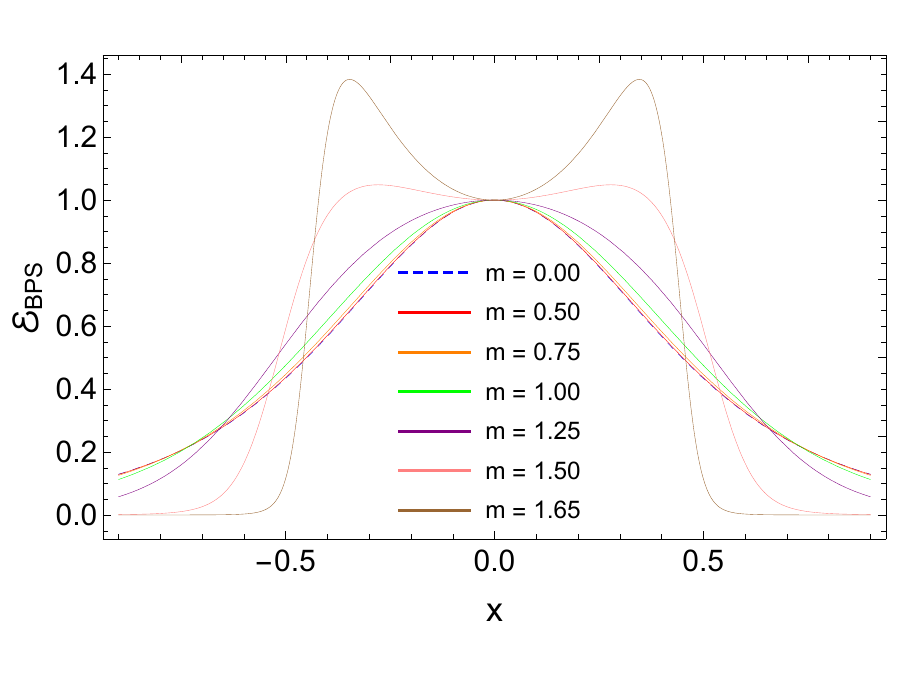}
\vspace{-0.3cm}
\caption{The BPS energy density concerning solutions of Eq. (\ref{Eq20}) with $\lambda=\nu=1$.} \label{fig4}
\end{figure}

Considering the numerical solutions presented in Figs. \ref{fig2}(a) and \ref{fig2}(b), we obtain BPS energy density exposed in Fig. \ref{fig4}. Likewise, one notes that the scalar field deforms into structures resembling the contracted kinks. We also observe the appearance of new critical energy points in the BPS energy density, i.e., for sufficiently large values of $m$, the value $\phi=0$ before was a local maximum turns a local minimum and two symmetric local maxima emerge.

\subsection{$\phi^{8}$ model: the case $f(\phi)=\left[\sinh(m\phi^2)\right]^{-\frac{1}{2z+1}}$}\label{SecIIb}

\begin{figure}[!ht]
\centering
\includegraphics[height=6cm,width=7cm]{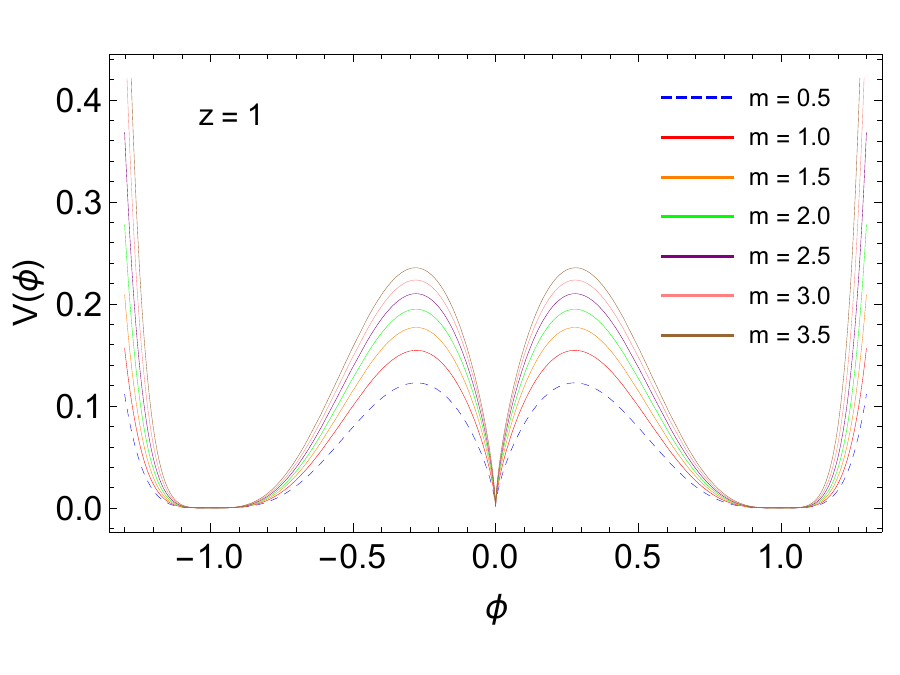} %
\includegraphics[height=6cm,width=7cm]{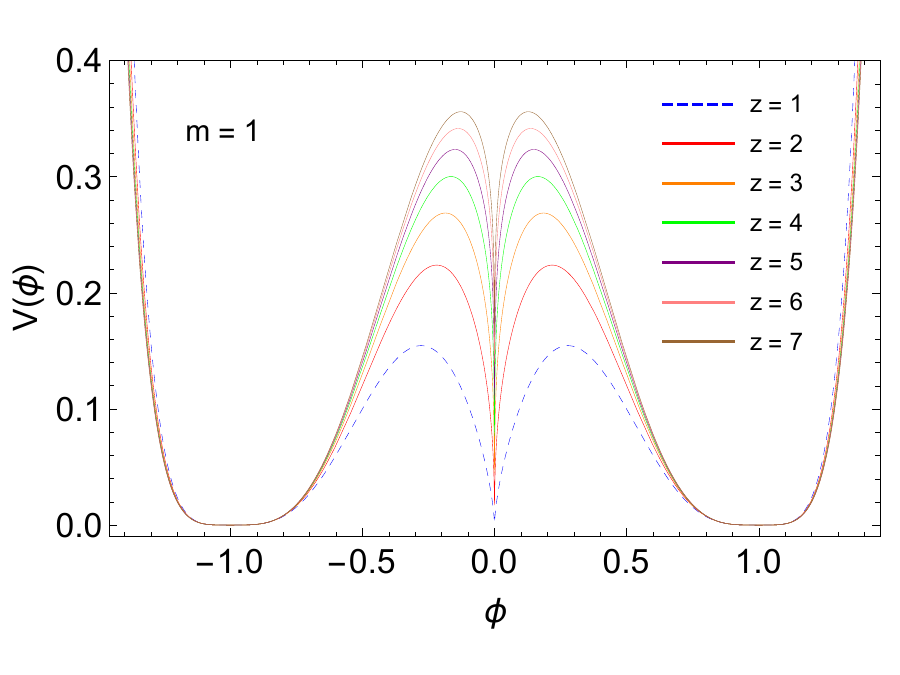}
\vspace{-0.2cm}

\hspace{0.8cm} (a) \hspace{6.5cm} (b) \hfill
\caption{Potential $V(\phi)$ vs. $\phi$ [Eq. (\ref{Eq25})]. The model: $f(\phi)=[\sinh(m \phi^{2})]^{-\frac{1}{2 z+1}}$ with $\lambda=\nu=1$.} \label{fig5}
\end{figure}

To study a second generalized $\phi^8$ model, one considers the generalizing function represented in Eq. (\ref{Eq18b}) and the superpotential (\ref{Eq14}). That way, the BPS potential reads as
\begin{align}\label{Eq25}
     V(\phi)= \frac{\lambda(\nu^{2}-\phi^{2})^4}{2} \left[\sinh(m\phi^2)\right]^{\frac{1}{2z+1}},
\end{align}
where $m>0$ and $z>1/2$ will be considered for our analysis. In this framework, the BPS equation is
\begin{align}\label{Eq26}
    \frac{d\phi}{dx}=\pm\sqrt{\lambda}(\nu^{2}-\phi^{2})^2 [\sinh(m\phi^2)]^{\frac{1}{2z+1}},
\end{align}
We are motivated to investigate this second model, characterized by the generalizing function $[\sinh(m \phi^{2})]^{-\frac{1}{2 z+1}}$, due to its ability to generate a potential's minimum at $\phi=0$ such as the Figs. \ref{fig5}(a) and \ref{fig5}(b) exhibit it.

\begin{figure}[!ht]
\centering
\includegraphics[height=6cm,width=7cm]{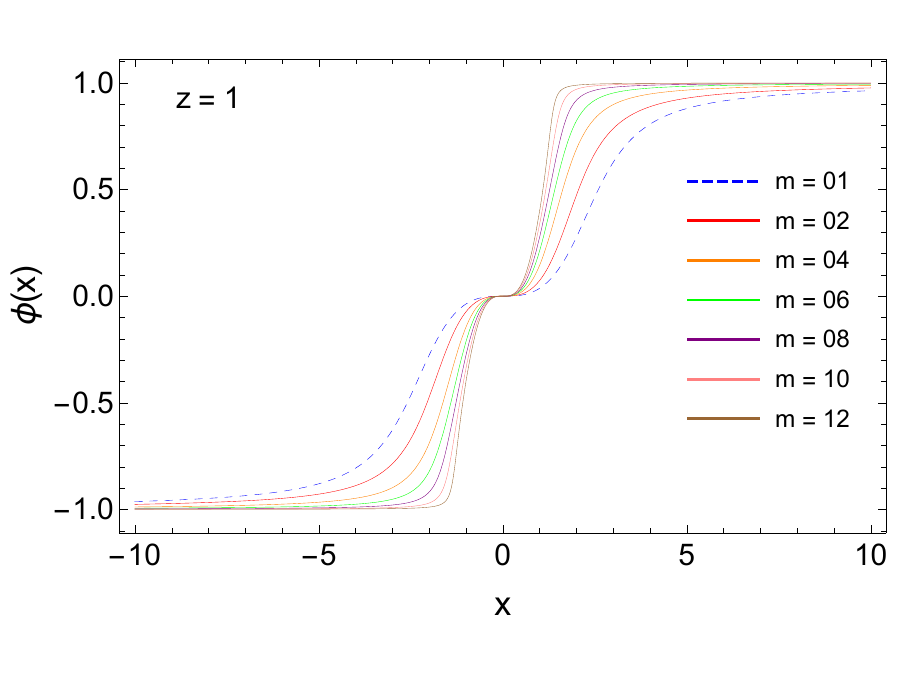}
\includegraphics[height=6cm,width=7cm]{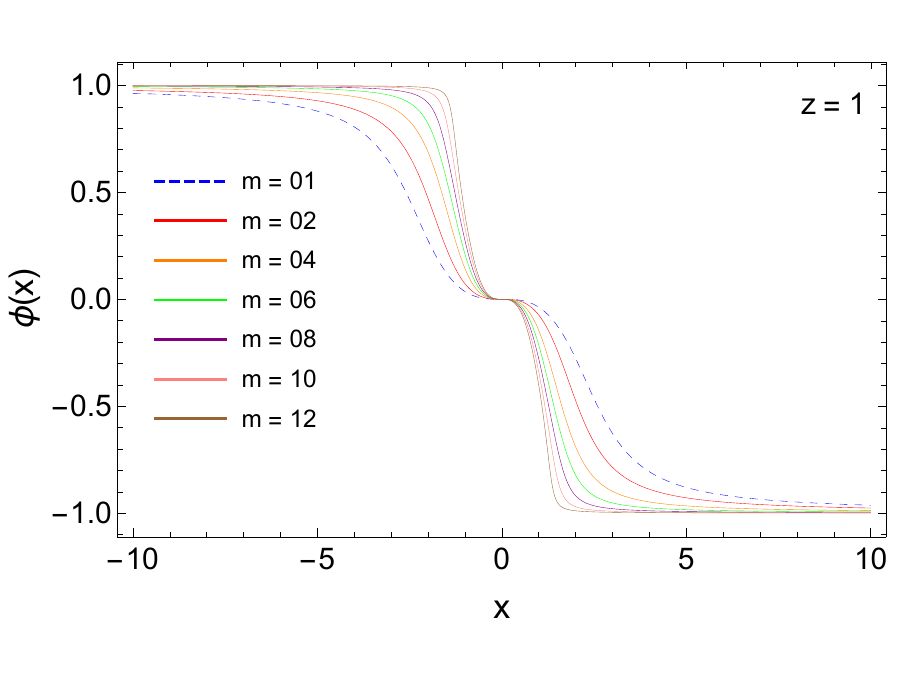} \vspace{-0.2cm}

\hspace{0.8cm} (a) \hspace{6.5cm} (b) \hfill\\
\includegraphics[height=6cm,width=7cm]{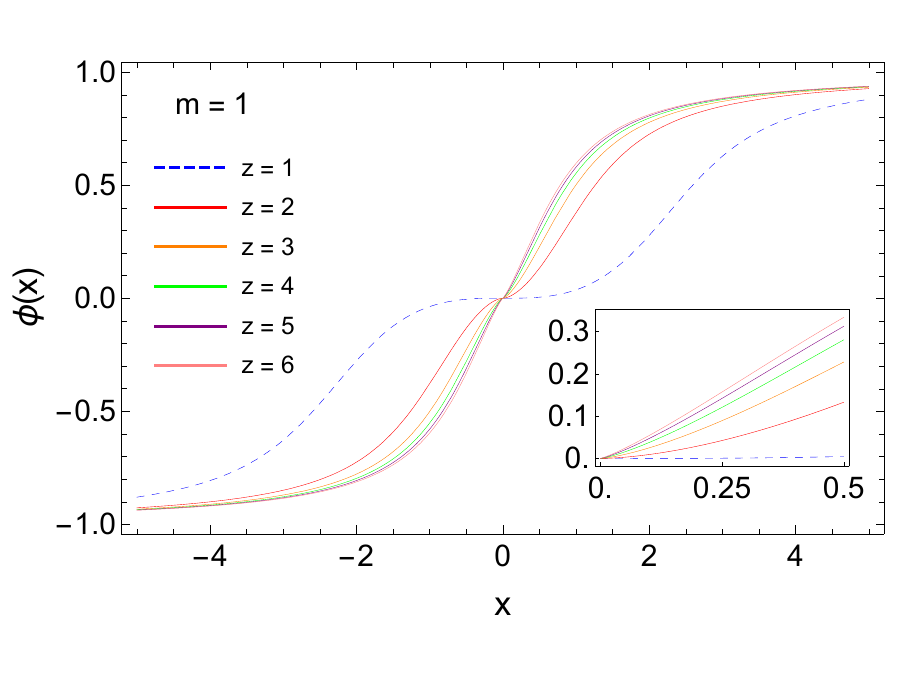}
\includegraphics[height=6cm,width=7cm]{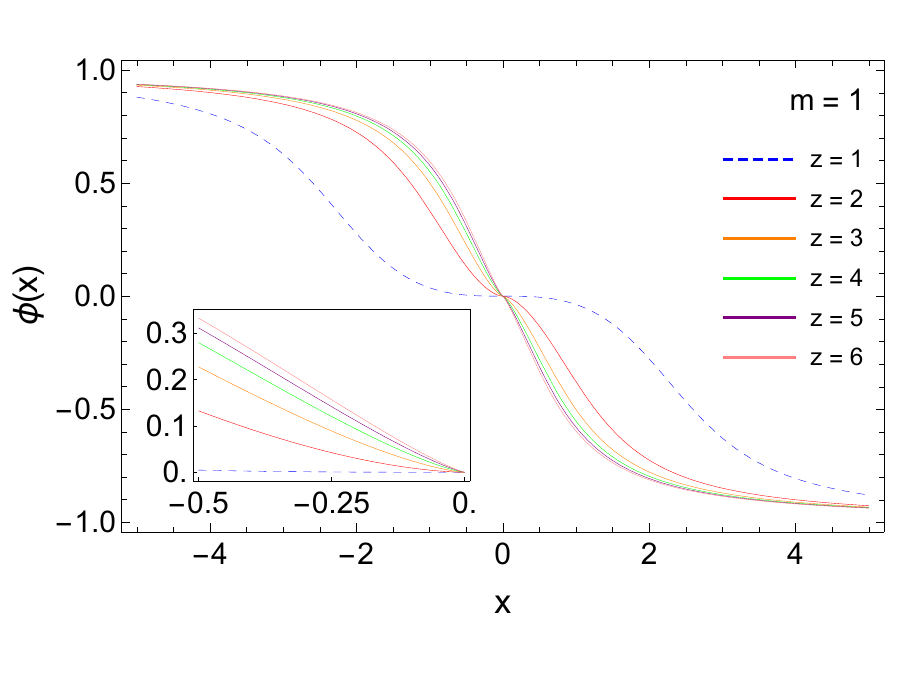}\vspace{-0.2cm}

\hspace{0.8cm} (c) \hspace{6.5cm} (d) \hfill
\caption{Plots $\phi(x)$ vs. $x$ keeping $\lambda=\nu=1$. The model $f(\phi)=[\sinh(m \phi^{2})]^{-\frac{1}{2 z+1}}$.} \label{fig6}
\end{figure}

The presence of $\phi=0$  minimum suggests the possibility of deformation of the original $\phi^8$ topological structures, whose consequence results in the formation of genuine double-kink configurations\footnote{One can find some projects in the literature by announcing double kinks' formation. For instance, in Ref. \cite{DBazeia00}, double kinks arise when analyzing the existence of topological configurations generated by a scalar field in a $(\text{D}+1)$-dimensional spacetime. In this framework, adopting a potential with minima critical points at $\pm 1$ and $0$, we note the arising of stable double-kink configurations. Furthermore, double kinks also emerge when considering the systematic quantum expansion of soliton solutions in a nonlinear field theory \cite{Christ}.}, as depicted by Figs. \ref{fig6}(a), \ref{fig6}(b), \ref{fig6}(c), and \ref{fig6}(d).

Furthermore, in the limit $z\to \infty$, we retrieve the $\phi^{8}$ whose self-dual equation is
\begin{align}\label{Eq28}
    \frac{d\phi}{dx}=\pm \sqrt{\lambda}(\nu^{2}-\phi^{2})^2,
\end{align}
which leads us to kink-like solutions, see Figs. \ref{fig6}(c) and \ref{fig6}(d), when the parameter $z$ increases. Therefore, one concludes that varying the $z$-parameter transforms double-kink-type solutions into kink-type ones. Moreover, increasing values of the $m$ parameter compactifies the format of soliton's profiles.

Studying the soliton behavior for $|x|\rightarrow\infty$ and near $x=0$ gives an enhanced understanding of the profiles' format. Such an analysis provides the following expressions,
\begin{eqnarray}
\phi (x)  &\approx &\mp v\pm \frac{1}{4v^{2}\sqrt{\lambda }[
\sinh( mv^{2})] ^{\frac{1}{2z+1}}}\frac{1}{\vert
x\vert },\;\; x\rightarrow -\infty, \quad \;    \\[0.2cm]
\phi (x)  &\approx &\pm \left( m\right) ^{\frac{1}{2z-1}}\left(
v^{4}\sqrt{\lambda }\frac{2z-1}{2z+1}x\right) ^{\frac{2z+1}{2z-1}},\;\; x\rightarrow 0,   \\[0.2cm]
\phi (x)  &\approx &\pm v\mp \frac{1}{4v^{2}\sqrt{\lambda }[\sinh( mv^{2})] ^{\frac{1}{2z+1}}}\frac{1}{\vert x\vert },\;\;x\rightarrow +\infty.
\end{eqnarray}
We again perceive that the profiles maintain the tail decaying as $|x|^{-1}$ such as the original $\phi^8$ solitons. On the other hand, the behavior at $x=0$ confirms that we have obtained genuine double-kink configurations, which transform into the kink solutions of the original $\phi^8$ model when $z\rightarrow +\infty$.

\begin{figure}[!ht]
\centering
\includegraphics[height=6cm,width=7cm]{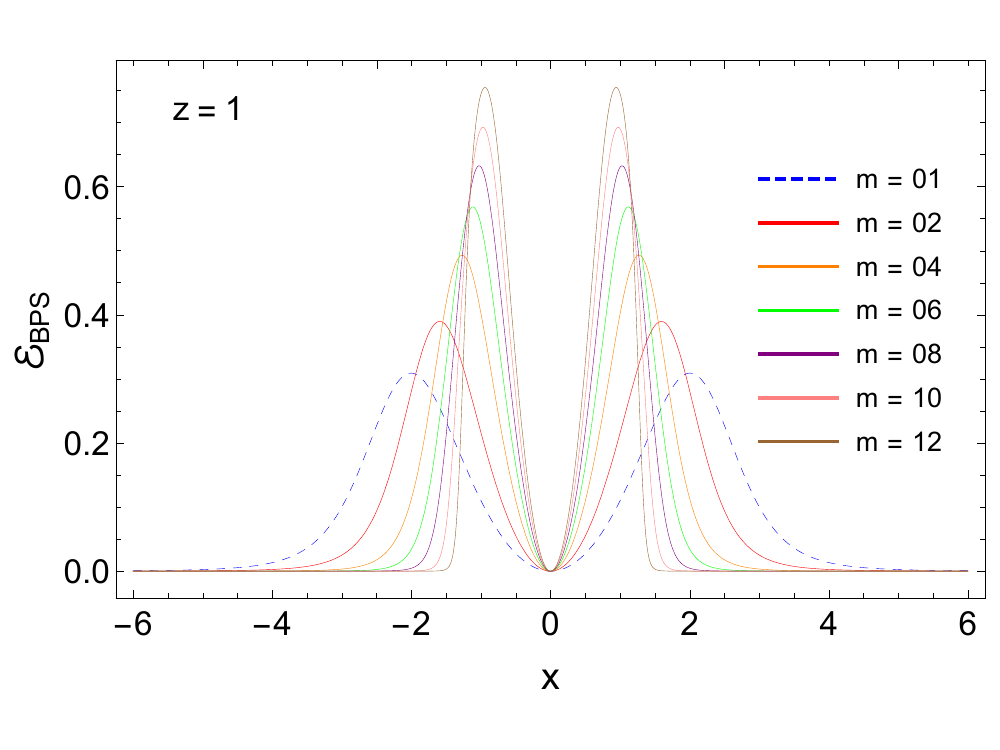}
\includegraphics[height=6cm,width=7cm]{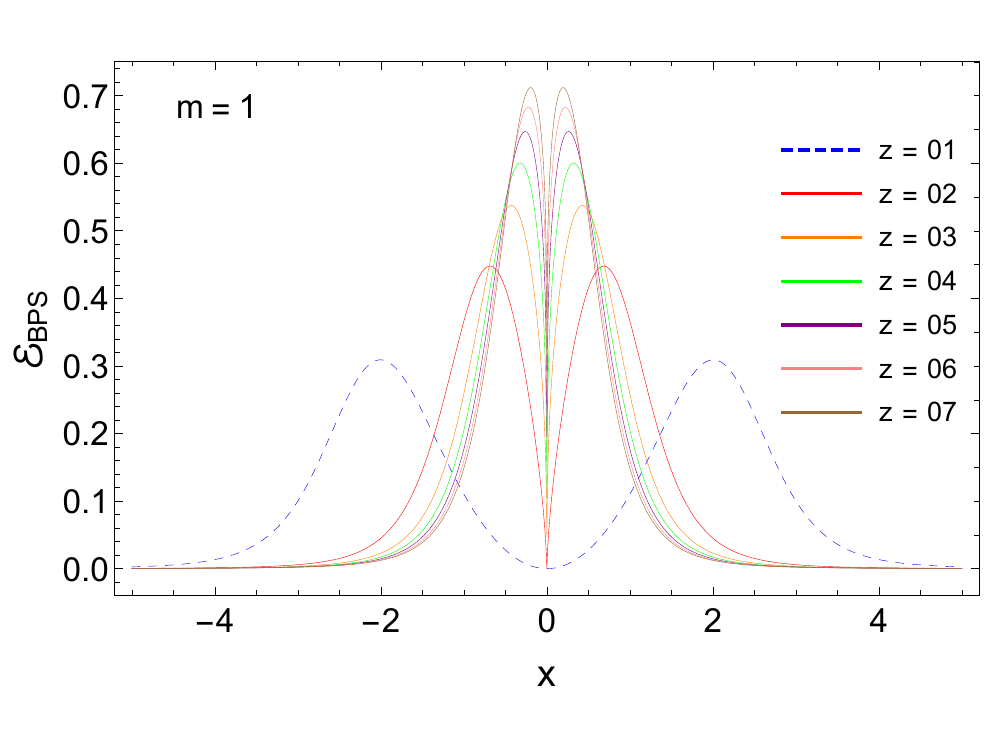}\vspace{-0.2cm}

\hspace{0.8cm} (a) \hspace{6.5cm} (b) \hfill\\
\caption{The BPS energy density when $f(\phi)=[\sinh(m \phi^{2})]^{-\frac{1}{2 z+1}}$.} \label{fig7}
\end{figure}

Furthermore, the BPS energy density given by
\begin{align}\label{Eq29}
    \mathcal{E}_{\textrm{BPS}}=\lambda(\nu^{2}-\phi^{2})^4 [\sinh(m\phi^2)]^{\frac{1}{2z+1}},
\end{align}
which is shown in Figs. \ref{fig7}(a) and \ref{fig7}(b). Notably, the analysis of the BPS energy density profiles proves the existence of double-kink solutions. This confirmation is due to the energy profile splitting into two symmetric parts around $x=0$. Naturally, by increasing the parameter $m$, the energy regions become more localized. Meanwhile, increasing the $z$ parameter, i.e., $z\to+\infty$, the BPS energy density becomes similar to kink-like configurations.

\subsection{$\phi^{4}$ model: the case $f(\phi)= \textrm{sech}^2(m^2 \phi) $} \label{SecIIc}

\begin{figure}[!ht]
\centering
\includegraphics[height=6cm,width=7cm]{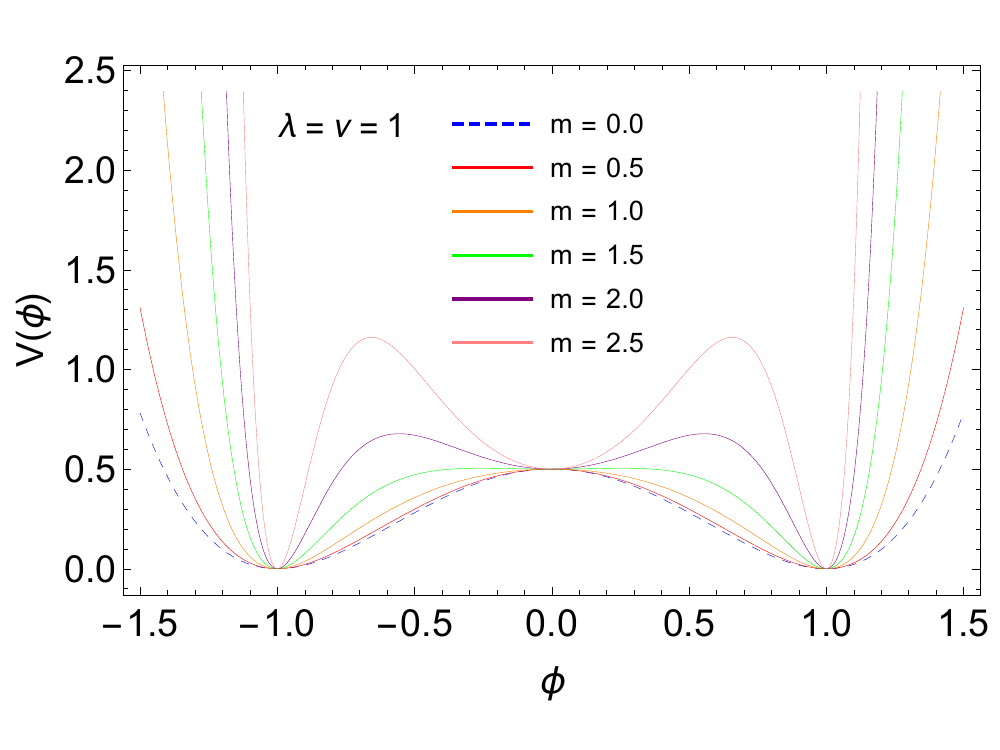}
\caption{Potential $V(\phi)$ vs. $\phi$ [Eq. (\ref{Eq30})] varying the parameter $m$.} \label{fig8}
\end{figure}

For the $\phi^{4}$ model with generalizing function Eq. (\ref{Eq18a}), one arrives to the BPS potential (\ref{Eq17}) being written as
\begin{align}\label{Eq30}
    V(\phi)=\lambda(\nu^{2}-\phi^{2})^2 \cosh(m^2\phi)^2,\quad  m\geq 0.
\end{align}
which has its profiles are depicted in Fig. \ref{fig8}. Moreover, it is clear that for $m=0$, one recovers the original $\phi^4$ theory.

\begin{figure}[!ht]
\centering
\includegraphics[height=6cm,width=7cm]{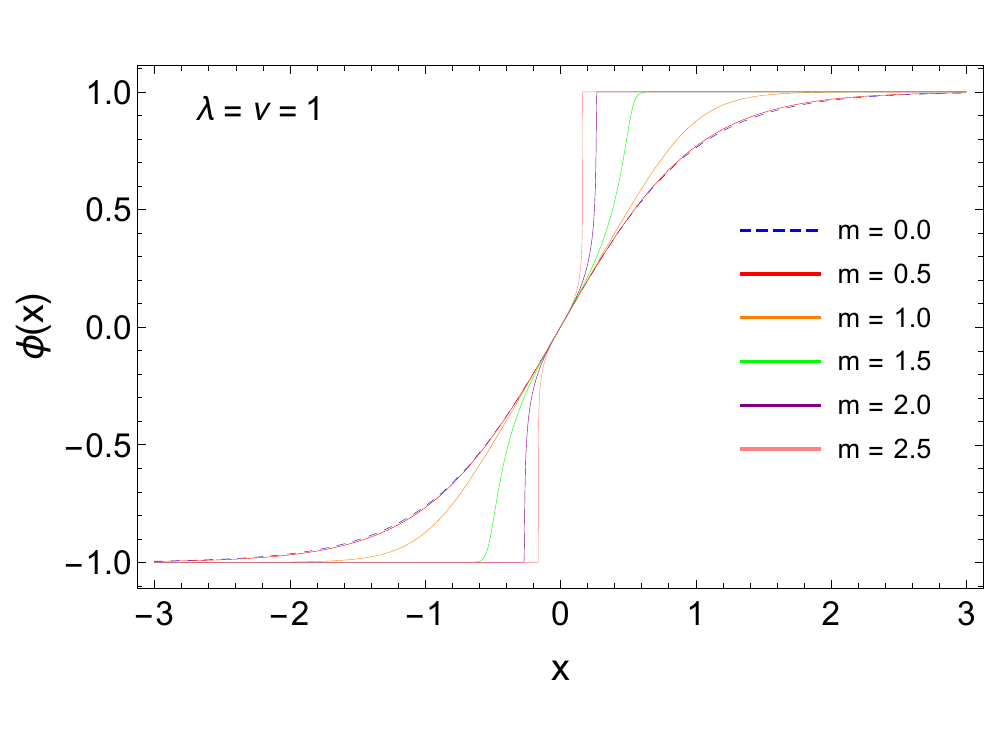}
\includegraphics[height=6cm,width=7cm]{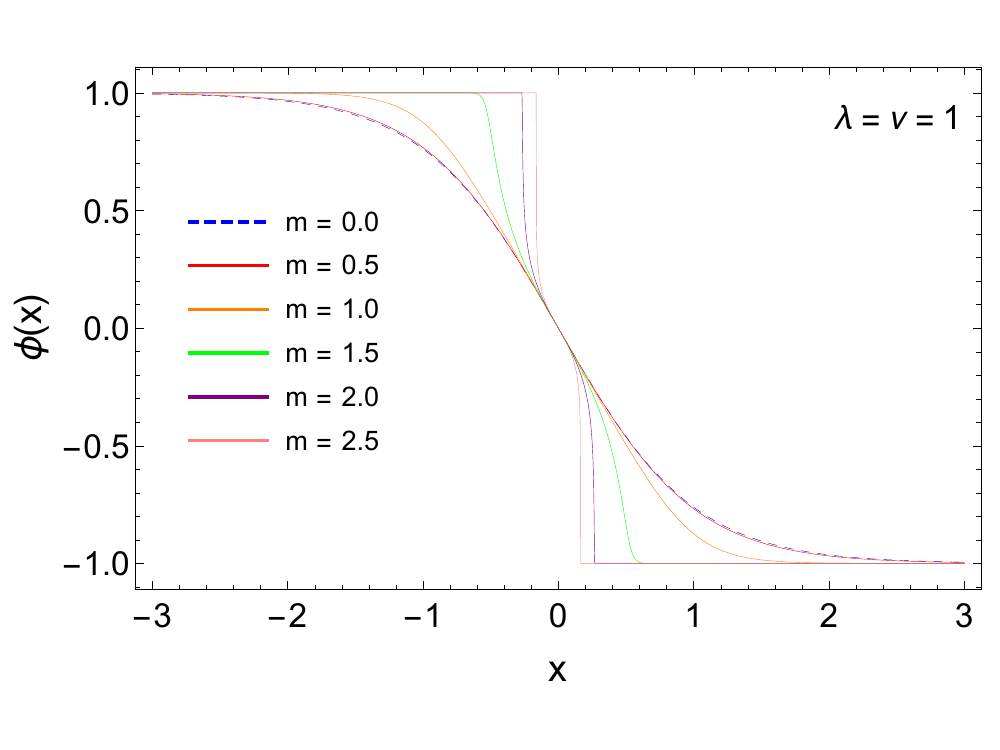}\vspace{-0.2cm}

\hspace{0.8cm} (a) \hspace{6.5cm} (b) \hfill\\
\caption{Plots $\phi(x)$ vs. $x$ keeping $\lambda=\nu=1$ [Eq. (\ref{Eq31})].} \label{fig9}
\end{figure}

Such as it happens in the $\phi^8$ case (see Fig. \ref{fig1}), owing to the contribution of the generalized function, for small values of $m$, the BPS potential looks like the original $\phi^{4}$ model. For increasing values of $m$, it gains a wider or less localized profile, while the local maximum at $\phi=0$ transforms into a local minimum. At the same time, the potential acquires two symmetric local maximums, which is related to the fact that solutions acquire a compacton-like format; see Figs. \ref{fig9}(a) and \ref{fig9}(b). Therefore, one concludes that the scalar field profiles resemble the structures of the generalized $\phi^{8}$ theory previously analyzed in Sec. \ref {SecIIa}.

In this case, the BPS equation is
\begin{align}\label{Eq31}
    \frac{d\phi}{dx}=\pm\sqrt{\lambda}(\nu^{2}-\phi^{2 })\cosh(m^2\phi)^2.
\end{align}

In the following, we present the field behavior in the asymptotic limit ($|x|\rightarrow\infty$) and at $x=0$,
\begin{eqnarray}
\phi (x)  &\approx &\mp v\pm C e^{-M \vert x \vert }, \;\; x\rightarrow -\infty\quad \;  \\[0.2cm]
\phi(x)  &\approx &\pm v^{2}\sqrt{\lambda }\,x, \;\; x\rightarrow 0  \\[0.2cm]
\phi (x) &\approx &\pm v\mp C e^{-M \vert x \vert },\;\; x\rightarrow -\infty,
\end{eqnarray}
where $C>0$  and $M$ represents the BPS mass of the solitons given by
\begin{equation}
M=2v\sqrt{\lambda }\cosh ^{2}\left( m^{2}v\right).
\end{equation}
The analysis indicates that the profiles retain the exponential-law decay of the original $\phi^4$, whereas the behavior at the origin strongly suggests that we only have kink/anti-kink configurations. Additionally, one compares the Figs. \ref{fig4} and \ref{fig9}, we observe that because of the tail exponential decay of the $\phi^4$ solitons, they develop a compact-like format more rapidly than the ones of the $\phi^8$ case for large values of $m$.

\begin{figure}[!ht]
\centering
\includegraphics[height=6cm,width=7cm]{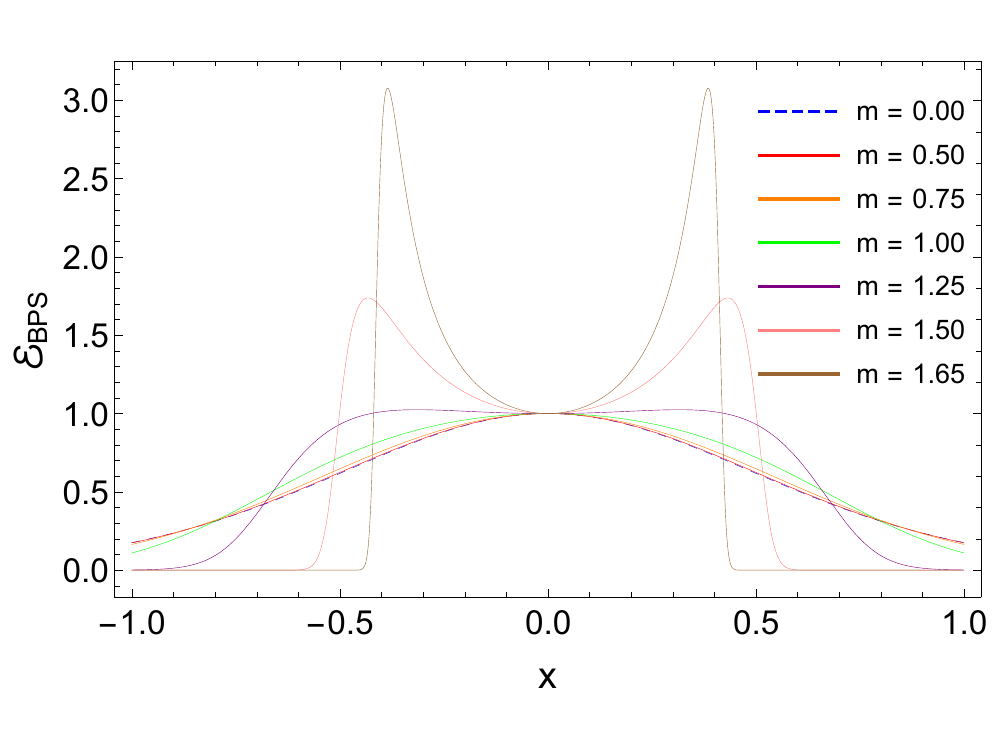}
\caption{The BPS energy density of the scalar field [Eq. (\ref{Eq33})].} \label{fig11}
\end{figure}

These field configurations carry a BPS energy expressed as
\begin{align}\label{Eq32}
    \mathrm{E}_{\mathrm{BPS}}=\pm\frac{4}{3}\sqrt{\lambda}\nu^3.
\end{align}
Besides, the BPS energy density given by
\begin{align}\label{Eq33}
    \mathcal{E}_{\textrm{BPS}}=\pm\lambda (\nu^{2}-\phi^{2})^2\cosh(m^2 \phi)^2,
\end{align}
has its profiles displayed in Fig. \ref{fig11}, where one notes that they are very similar to the ones of the $\phi^8$ case exhibited in Fig. \ref{fig4}.

\subsection{$\phi^{4}$ model: the case $f(\phi)=[\sinh(m \phi^{2z})]^{-\frac{1}{2 z+1}}$\label{SecIId}}

\begin{figure}[!ht]
\centering
\includegraphics[height=6cm,width=7cm]{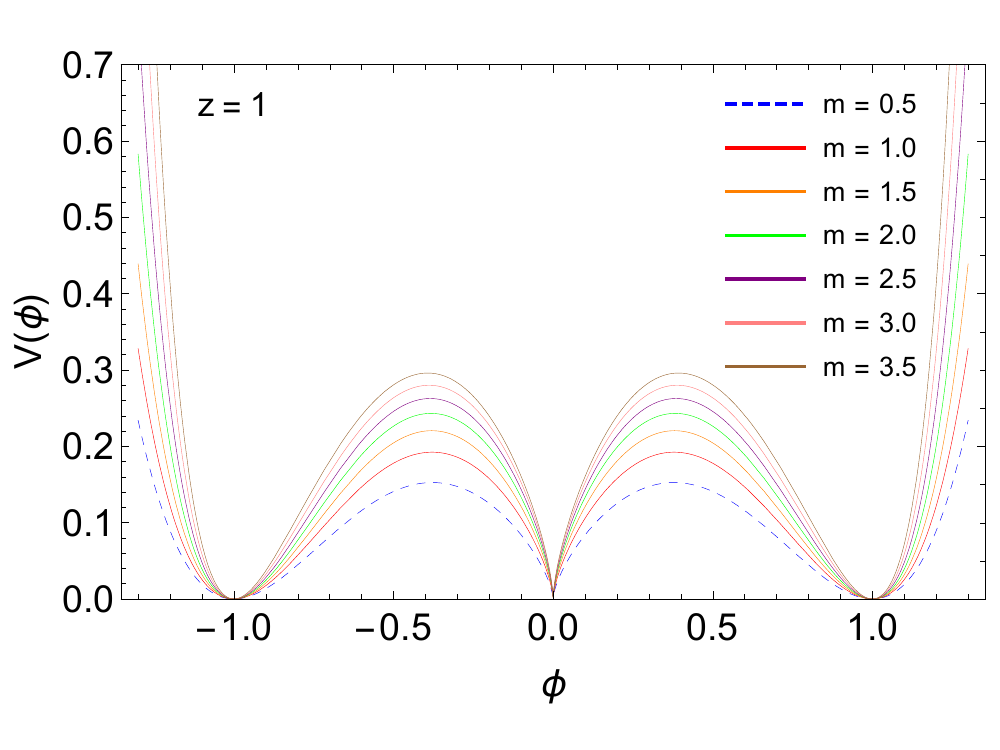}
\includegraphics[height=6cm,width=7cm]{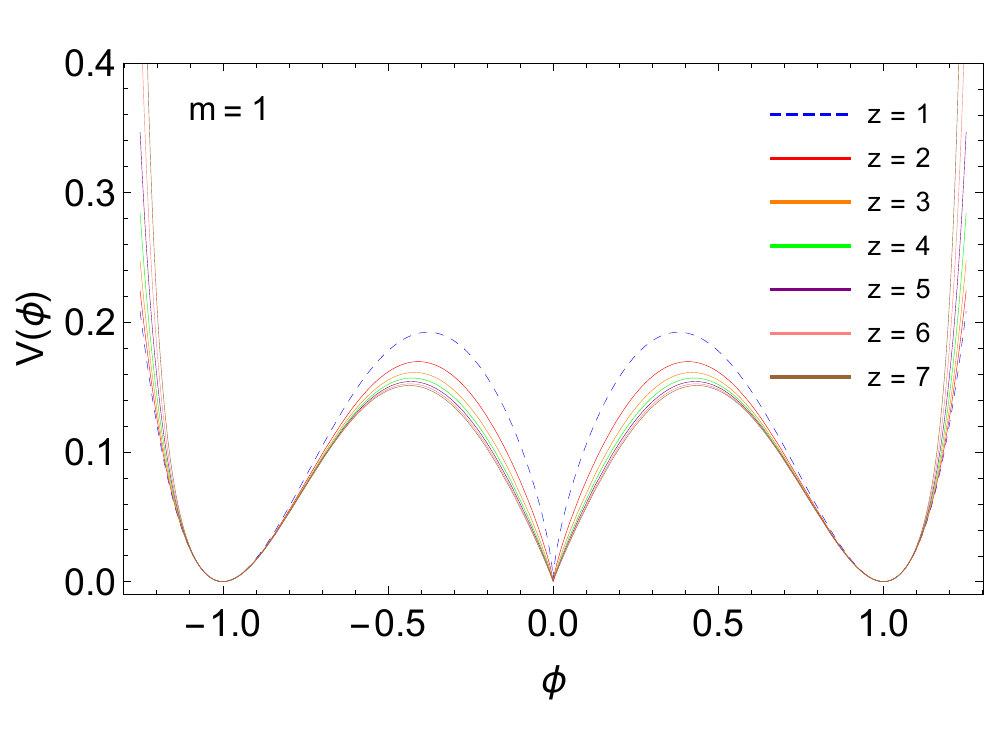}\vspace{-0.2cm}

\hspace{0.8cm} (a) \hspace{6.5cm} (b) \hfill
\caption{Potential $V(\phi)$ vs. $\phi$ [Eq. (\ref{Eq34})]. (a) Varying the parameter $m$ and keeping $z=1$. (b) Varying the parameter $z$ and keeping $m=1$.} \label{fig12}
\end{figure}

Now, let us analyze the final case by adopting the generalizing function given by Eq. (\ref{Eq18c}) in the modified $\phi^{4}$-like theory. In this case, the BPS potential is
\begin{align}\label{Eq34}
    V(\phi)=\frac{\lambda}{2}(\nu^{2}-\phi^{2})^2 [\sinh(m\phi^{2z})]^{\frac{1}{2z+1}},
\end{align}
where we will consider  $m>0$ and $z\geq 0$. We note that for $z=0$, the potential becomes that of the original $\phi^4$ model. For $z>0$, one notes that because of the hyperbolic sine function, we again have three vacuum values $\phi=-v, 0, v$. Consequently, the emergence of genuine double-kink configurations is guaranteed. For fixed $z$ and increasing values of $m$, the amplitude of the potential barrier around $\phi=0$ becomes more pronounced; meantime, for fixed $m$ and growing values of $z$, such amplitude diminishes. For further details, the Figs. \ref{fig12}(a) and \ref{fig12}(b) illustrate the profiles of the BPS potential.

In this case, the BPS equation is
\begin{align}\label{Eq35}
    \frac{d\phi}{dz}=\pm\sqrt{\lambda}(\nu^{2}-\phi^{2}) [\sinh(m\phi^{2z})]^{\frac{1}{2z+1}}.
\end{align}

\begin{figure}[!ht]
\centering
\includegraphics[height=6cm,width=7cm]{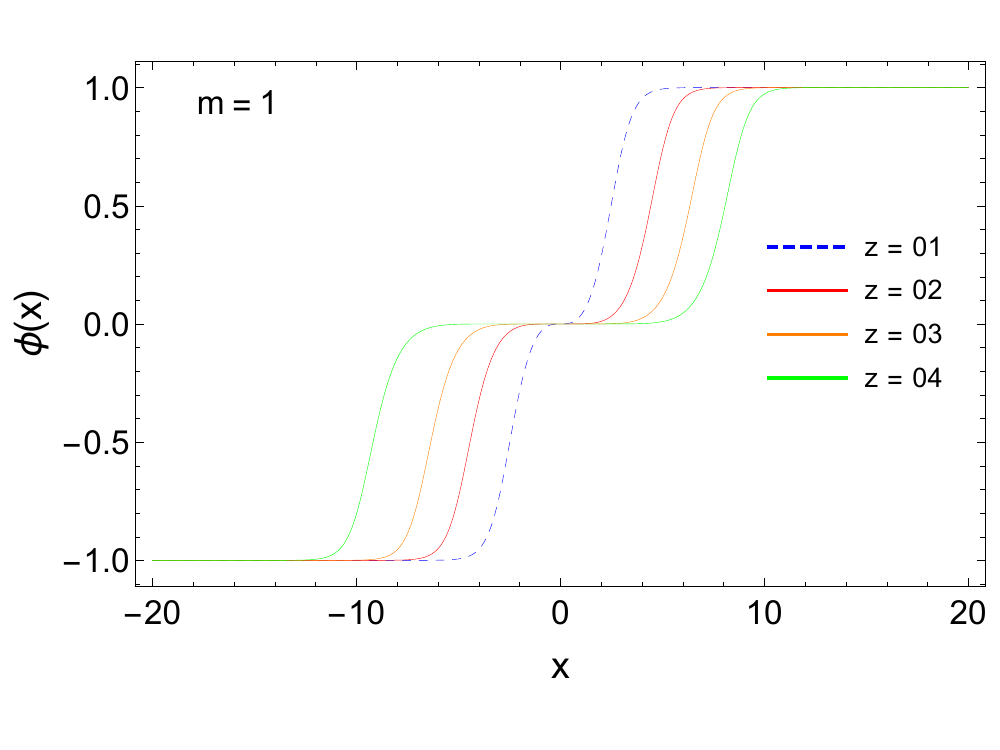}
\includegraphics[height=6cm,width=7cm]{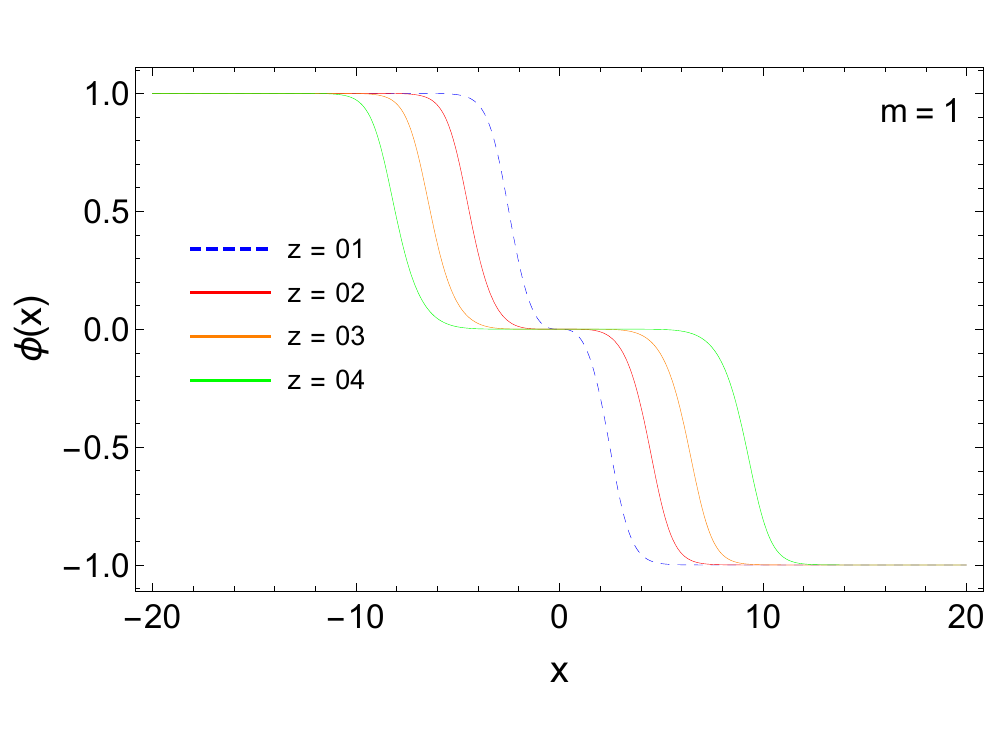}\vspace{-0.2cm}

\hspace{0.8cm} (a) \hspace{6.5cm} (b) \hfill\\
\includegraphics[height=6cm,width=7cm]{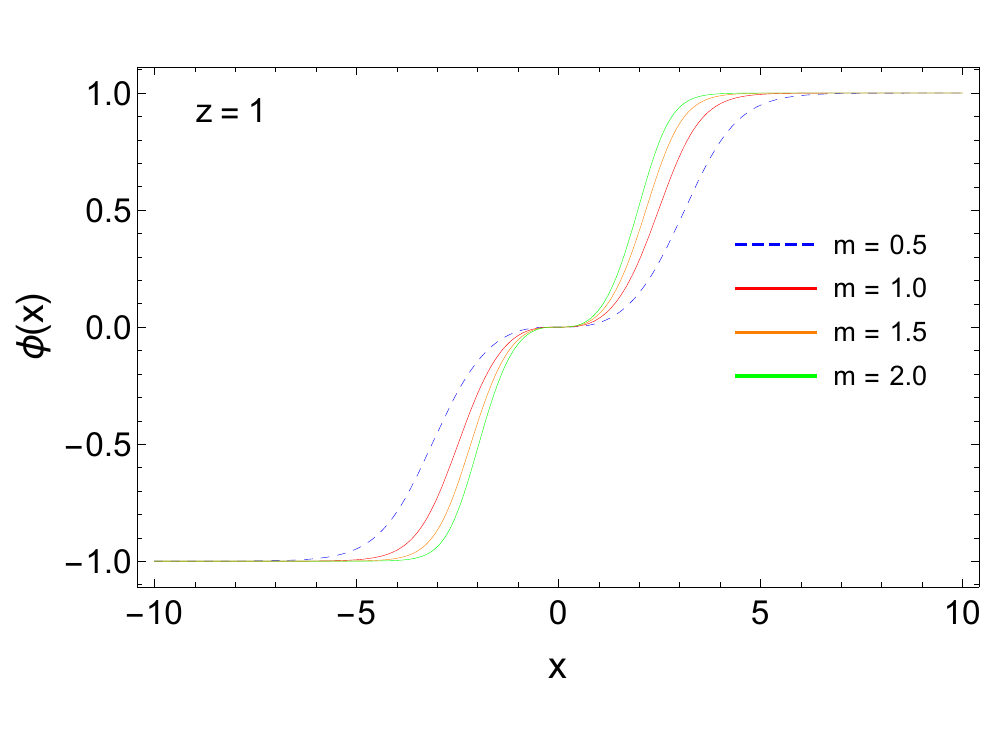}
\includegraphics[height=6cm,width=7cm]{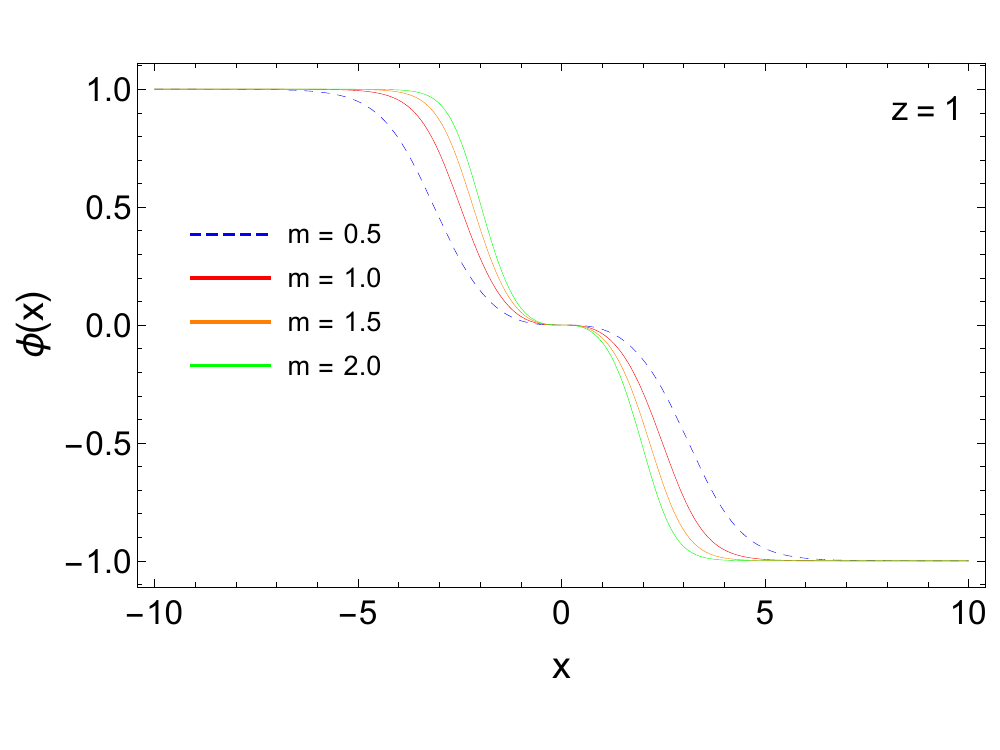} \vspace{-0.2cm}

\hspace{0.8cm} (c) \hspace{6.5cm} (d) \hfill
\caption{Solutions of the scalar field [Eq. (\ref{Eq35})]. Plots of $\phi(x)$ vs. $x$ adopting the generalizing function $f(\phi)=[\sinh(m \phi^{2z})]^{-\frac{1}{2 z+1}}$.} \label{fig13}
\end{figure}

Likewise, we also write the behavior of the solutions for $x\rightarrow\pm\infty$ and around $x=0$,
\begin{eqnarray}
\phi(x)  &\approx &\mp v\pm \mathcal{C} e^{-\mathcal{M}\vert x\vert },\;\; x\rightarrow -\infty,\quad\;  \\[0.2cm]
\phi(x)   &\approx &\pm m\left( v^{2}\sqrt{\lambda }\frac{x}{2z+1%
}\right) ^{2z+1},\;\;x\rightarrow 0,\\[0.2cm]
\phi(x)  &\approx &\pm v\mp \mathcal{C}
e^{-\mathcal{M}\vert x\vert },\;\; x\rightarrow +\infty,
\end{eqnarray}
where $\mathcal{C}>0$  and $\mathcal{M}$ is the BPS mass of the solitons,
\begin{equation}
\mathcal{M}=2v\sqrt{\lambda }\left[ \sinh \!\left( mv^{2z}\right) \right] ^{\frac{1}{2z+1}}.
\end{equation}
We observe that the asymptotic behavior, i.e., how the field comports near the vacuum values, remains similar to one in the original $\phi^4$ model. On the other hand, for $z>0$, the behavior is modified because the potential acquires the minimum at $\phi=0$, guaranteeing the rising of genuine double-kink solitons, see Figs. \ref{fig13}[(a)-(d). Moreover, for $z=0$, we obtain a behavior similar to that of the original $\phi^4$ kinks.

Figures \ref{fig13}[(a)-(d)] show the influence of the $z$ and $m$ parameters on the profiles. By maintaining $m$ fixed, the Figs. \ref{fig13}[(a)-(b)] show that the plateau around the origin becomes wider for increasing values of $z$, so the width of the soliton grows. Otherwise, Figs. \ref{fig13}[(c)-(d)] exhibit that for fixed $z$ and growing values of $m$ the solutions get a more compact-like format.

\begin{figure}[!ht]
\centering
\includegraphics[height=6cm,width=7cm]{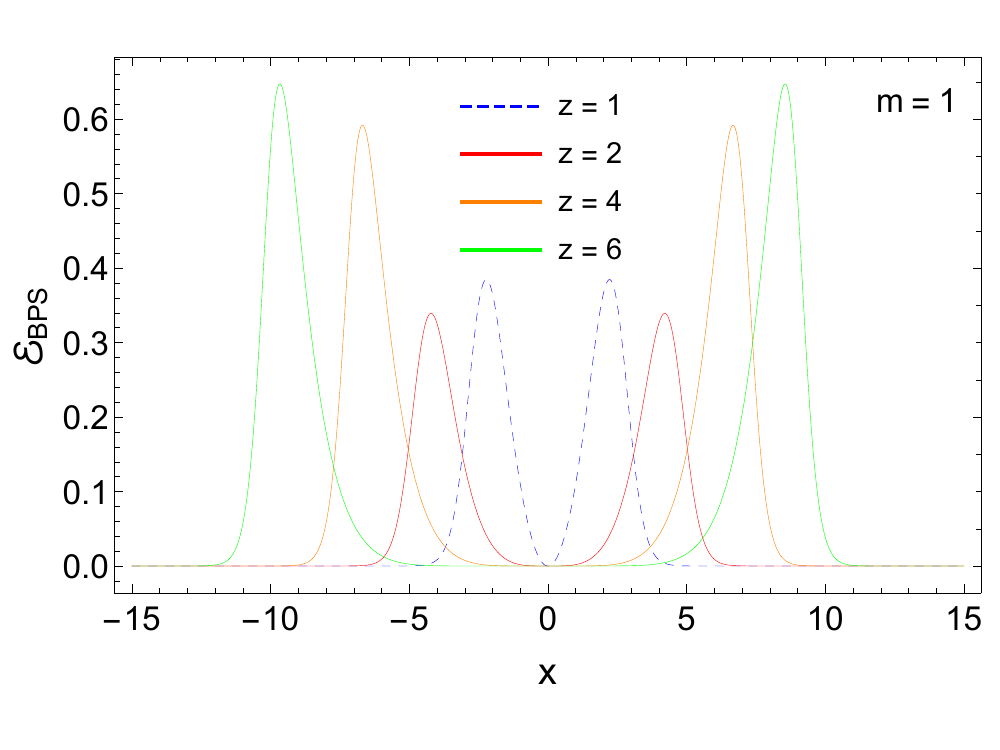} %
\includegraphics[height=6cm,width=7cm]{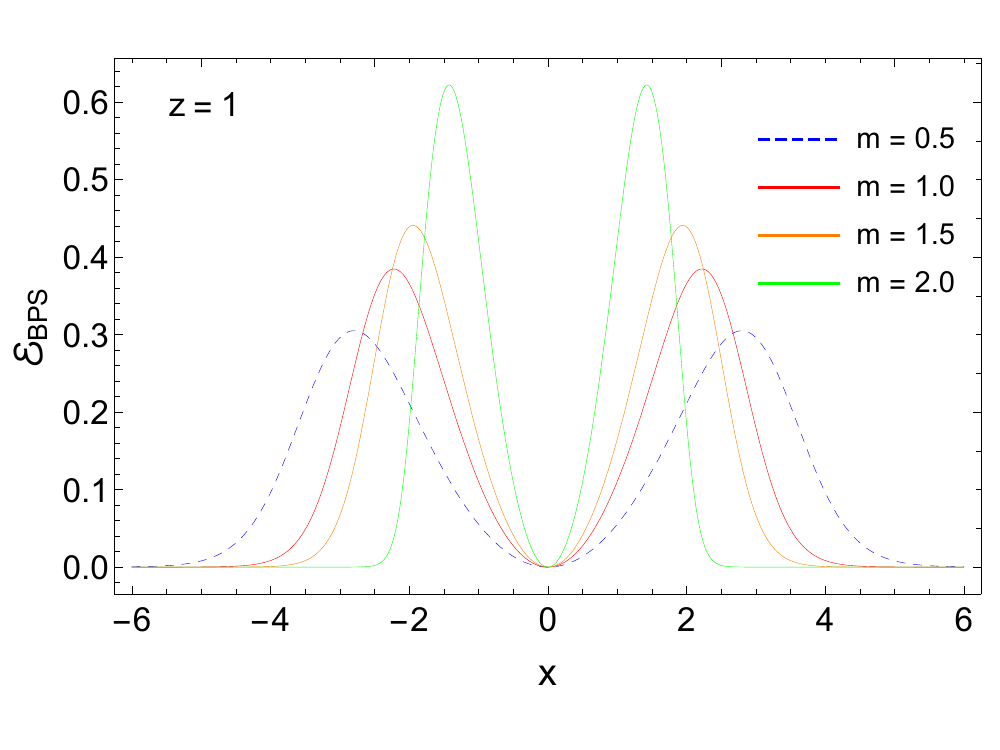}  \vspace{-0.2cm}

\hspace{0.8cm} (a) \hspace{6.5cm} (b) \hfill
\caption{The BPS energy density of the scalar field [Eq. (\ref{Eq36})].} \label{fig14}
\end{figure}

Furthermore, it is necessary to highlight that these configurations possess the BPS energy as announced in Eq. (\ref{Eq32}), and the respective BPS energy density will be
\begin{align}\label{Eq36}
    \mathcal{E}_{\textrm{BPS}}=\pm\lambda(\nu^{2}-\phi^{2})^2 [\sinh(m\phi^{2z})]^{\frac{1}{2z+1}},
\end{align}
whose numerical solutions are exposed in Figs. \ref{fig14}[(a)-(b)].

\section{Final remarks}\label{SecIV}

Throughout this study, we examined both the  $\phi^{4}$ and $\phi^{8}$ theories into a two-dimensional flat space-time. By incorporating noncanonical contributions from the kinetic term, i.e., $\frac{1}{2}f(\phi) \partial^\mu\phi \partial_\mu\phi$, one noted that the noncanonical model permits new classes of solutions resembling kink- and double-kink-type configurations. Furthermore, one observed that we can also compact these new configurations with appropriate adjustments of the generalizing function's parameters.

The theories $\phi^{8}$ and $\phi^{4}$ utilized were, respectively, described by the interactions
\begin{align}
    V(\phi)=\frac{\lambda(\nu^{2}-\phi^{2})^4}{2f(\phi)},\;\;\text{and} \quad V(\phi)=\frac{\lambda(\nu^{2}-\phi^{2})^2}{2f(\phi)}.
\end{align}
In this scenario, the chosen generalizing functions are $[\cosh(m^2\phi)]^{-2}$, $[\sinh(m\phi^2)]^{-\frac{1}{2z+1}}$, and $[\sinh(m\phi^{2z})]^{-\frac{1}{2z+1}}$. Among these choices, the one with the hyperbolic sine induces the appearance of a potential's minimum at $\phi=0$. Thus, true double-kink-type solutions are achieved in both generalized $\phi^{8}$ and $\phi^{4}$ models. Conversely, the one containing the hyperbolic cosine causes the formation of kink-type solitons. Thereby, in all the cases, the minimal energy configurations geometrically deformed arise, announcing new classes of scalar field solutions.

One highlights that the selected generalized functions are also suitable due to their behavior relative to the parameters on which they depend. For instance, considering the generalizing function $f(\phi)=\textrm{sech}^2 (m^2\phi)$, we reached the usual $\phi^{4}$ and $\phi^{8}$  theories when $m\to 0$. Moreover, in both generalized models, the field's profiles for $x=0$ and when $|x|\rightarrow \infty$ possess behavior similar to the ones obtained in the usual models, it happens because the generalizing function (\ref{Eq18a}) is a well-behaved function $\forall\, x \in \mathds{R}$. The usual theories are recovered in the case $f(\phi)=[\sinh(m\phi^2)]^{-\frac{1}{2z+1}}$ in the limit $z\to\infty$. On the other hand, for $f(\phi)=[\sinh(m\phi^{2z}) ]^{-\frac{1}{2z+1}}$ we recuperate them when $z\to 0$, while when $z\to\infty$ one obtains a polynomial-like generalizing theory, i.e.,   $f(\phi)=|\phi|^{-1}$. Furthermore, both generalized functions involving the hyperbolic sine function are singular at $x=0$ (when $z>0$), being it the true cause for the behavior changing of the soliton profile around the origin, an effect of the potential's minimum at $\phi=0$. Consequently, it is also related to the rise of the genuine double-kink-type solitons. Lastly, it is worth highlighting that the increasing values of the $m$ parameter shrink  the field solutions, developing a compact-like format in all the generalized models studied here. Moreover, the hyperbolic generalizing functions have allowed us to obtain double-domain walls, as shown in Figs.  \ref{fig7} and \ref{fig14} using the $\phi^8$ and $\phi^4$ models with unusual kinematics, respectively.

\section{Acknowledgment}

The authors express their gratitude to FAPEMA and CNPq (Brazilian research agencies) for their invaluable financial support. F. C. E. L. is supported by FAPEMA BPD-05892/23. R. C. acknowledges the support from the grants CNPq/312155/2023-9, FAPEMA/UNIVERSAL-00812/19, and FAPEMA/APP-12299/22. C. A. S. A. is supported by CNPq 309553/2021-0 (CNPq/Produtividade) and project 200387/2023-5. Furthermore, C. A. S. A. acknowledges the Department of Theoretical Physics \& IFIC from the University of Valencia for their warm hospitality.

\section*{Appendix A - Rescaling the non-canonical theory\label{AppA}}

Let us consider the non-canonical model described by the action \eqref{Eq2}, i.e.,
\begin{align}
    \label{AEq1}
        S=\int\, d^2 x\, \bigg[\frac{1}{2}f(\phi)\,\partial^\mu \phi\,\partial_\mu\phi-V(\phi)\bigg],
\end{align}
where $f(\phi)$ is a positive definite function to preserve the sign of the Lagrangian density and avoid the presence of ghosts, negative energy densities, and negative energies.  Now, let us employ an approach resembling the one announced by Losano et al. \cite{LLosano}. Thus, we define the scalar field $\Phi$ as 
\begin{align}\label{AEq2}
    \Phi\equiv\int^{\phi}\, d\tilde{\phi}\, \sqrt{f(\tilde{\phi})}= g(\phi),
\end{align}
such that we obtain
\begin{align}
   g_\phi=\sqrt{f(\phi)}.
\end{align}
This way, we achieve
\begin{align}
    \label{AEq3}
\partial_\mu\Phi=g_\phi\,\partial_\mu\phi=\sqrt{f(\phi)}\partial_\mu\phi.
\end{align}

By regarding Eq. \eqref{AEq3}, one obtains the relation
\begin{align}
    \label{AEq4}
    \partial_\mu\Phi\partial^\mu\Phi=f(\phi)\,\partial_\mu\phi\,\partial^\mu\phi,
\end{align}
allowing us to rewrite the action (\ref{AEq1}) in the following canonical form
\begin{align}\label{AEq5}
    S_\text{can}=\int\, dx^2\, \left[\frac{1}{2}\partial_\mu\Phi\,\partial^\mu\Phi-V(g^{-1}(\Phi)) \right].
\end{align}

Once performing the field transformation $\phi\to \Phi$, the noncanonical kinetic term turns the canonical one, but the potential defined as $V(\phi)$ becomes deformed to be $V(g^{-1}(\Phi))$. In principle, both models will describe the equivalent physical solutions. The literature shows one can deform the field configurations using two approaches: the first involves modifying the potential \cite{Def1,Def2}, and the second deforms the field configurations by implementing the generalizing function in the kinetic term. Nevertheless, the generalized approach has gained more popularity, as evidenced by Refs. \cite{Lima,DBazeia1,DBazeia2,DBazeia3}. That is because the generalizing function $f(\phi)$ adjusts the contribution of the kinetic term and is related to the potential when the field energy is saturated, thereby altering the behavior of the potential and deforming the structures; see Eq. \eqref{Eq8}. In light of these observations, we have chosen the approach based on the generalizing function to investigate new classes of topological structures.

\end{document}